\documentclass[conference]{IEEEtran}
\makeatletter
\renewcommand{\@fnsymbol}[1]{\@arabic{#1}}
\makeatother

\IEEEoverridecommandlockouts

\usepackage{fancyhdr} 

\usepackage{lipsum}
\usepackage[utf8]{inputenc}
\usepackage[T1]{fontenc}

\usepackage{listings}
\usepackage{xcolor}

\definecolor{codegreen}{rgb}{0,0.6,0}
\definecolor{codegray}{rgb}{0.5,0.5,0.5}
\definecolor{codepurple}{rgb}{0.58,0,0.82}
\definecolor{backcolour}{rgb}{0.95,0.95,0.92}

\lstdefinestyle{mystyle}{
    backgroundcolor=\color{backcolour},   
    commentstyle=\color{codegreen},
    keywordstyle=\color{magenta},
    numberstyle=\tiny\color{codegray},
    stringstyle=\color{codepurple},
    basicstyle=\ttfamily\footnotesize,
    breakatwhitespace=false,         
    breaklines=true,                 
    captionpos=b,                    
    keepspaces=true,                 
    numbers=left,                    
    numbersep=5pt,                  
    showspaces=false,                
    showstringspaces=false,
    showtabs=false,                  
    tabsize=2
}

\usepackage{graphicx} 
\usepackage{subfig}
\usepackage{nccmath}
\usepackage{color}
\usepackage{cite}
\usepackage{acro}
\usepackage{scalerel}
\usepackage{tikz}
\usetikzlibrary{svg.path}

\usepackage[absolute,overlay]{textpos} 

\usepackage{cite}
\usepackage{amsmath,amssymb,amsfonts}
\usepackage{algorithmic}
\usepackage{graphicx}
\usepackage{textcomp}
\usepackage{booktabs}
\usepackage{mathtools}
\usepackage{array}
\usepackage{multirow}
\usepackage{xcolor}
\usepackage{listings}
\def\BibTeX{{\rm B\kern-.05em{\sc i\kern-.025em b}\kern-.08em
    T\kern-.1667em\lower.7ex\hbox{E}\kern-.125emX}}

\usepackage{tabularx,booktabs}

\usepackage{tabularx,booktabs,array, multirow}
\newcolumntype{Y}{>{\raggedright\arraybackslash}X}

\usepackage{orcidlink}

\usepackage{booktabs,multirow,tabularx,array}
\newcolumntype{Y}{>{\raggedright\arraybackslash}X}

\definecolor{orcidlogocol}{HTML}{A6CE39}
\tikzset{
  orcidlogo/.pic={
    \fill[orcidlogocol] svg{M256,128c0,70.7-57.3,128-128,128C57.3,256,0,198.7,0,128C0,57.3,57.3,0,128,0C198.7,0,256,57.3,256,128z};
    \fill[white] svg{M86.3,186.2H70.9V79.1h15.4v48.4V186.2z}
                 svg{M108.9,79.1h41.6c39.6,0,57,28.3,57,53.6c0,27.5-21.5,53.6-56.8,53.6h-41.8V79.1z M124.3,172.4h24.5c34.9,0,42.9-26.5,42.9-39.7c0-21.5-13.7-39.7-43.7-39.7h-23.7V172.4z}
                 svg{M88.7,56.8c0,5.5-4.5,10.1-10.1,10.1c-5.6,0-10.1-4.6-10.1-10.1c0-5.6,4.5-10.1,10.1-10.1C84.2,46.7,88.7,51.3,88.7,56.8z};
  }
}

\newcommand\orcidicon[1]{\href{https://orcid.org/#1}{\mbox{\scalerel*{
\begin{tikzpicture}[yscale=-1,transform shape]
\pic{orcidlogo};
\end{tikzpicture}
}{|}}}}

\usepackage{xcolor}
\usepackage{soul}

\usepackage{hyperref} 

\begin{document}

\renewcommand{\thefootnote}{\arabic{footnote}}
\setcounter{footnote}{0}

\title{Barriers to Integrating Low-Power IoT in Engineering Education: A Survey of the Literature
}

\author{
    V. Sanchez Padilla\textsuperscript{1,2},~\textit{Senior Member,~IEEE},  
    Albert Espinal\textsuperscript{3},~\textit{Member,~IEEE}, \\ Jose Cordova-Garcia\textsuperscript{3},~\textit{Member,~IEEE},
    Lisa Schibelius\textsuperscript{1} \\
    \textsuperscript{1}Dept. of Engineering Education, Virginia Polytechnic Institute and State University, Blacksburg, VA 24060, USA \\ 
    \textsuperscript{2}College of Engineering, Universidad ECOTEC, Samborondón, EC092302, Ecuador \\ 
    \textsuperscript{3}Dept. of Telematics Engineering, Escuela Superior Politécnica del Litoral, Guayaquil, EC090902, Ecuador \\
    \{vsanchez, lisaschib\}@vt.edu, \{aespinal, jecordov\}@espol.edu.ec


}

\maketitle


\begin{textblock*}{\textwidth}(15mm,5mm) 
\centering
\footnotesize © 2025 IEEE. PREPRINT. This is the authors' version of the work accepted for presentation at 2025 IEEE 16th Annual Information Technology, Electronics and Mobile Communication Conference (IEMCON), Berkeley, CA, USA. 

\end{textblock*}

\begin{abstract}
Low-power Internet of Things (IoT) technologies are becoming increasingly important in engineering education as a tool to help students connect theory to real applications. However, many institutions face barriers that slow down their adoption in courses and labs. This paper reviews recent studies to understand these barriers and organizes them into three groups: technical, organizational, and curricular/pedagogical. Technical barriers include energy management, scalability, and integration issues. Organizational barriers are related to cost, planning, and the need for trained staff. Curricular and pedagogical barriers include gaps in student readiness, limited lab time, and platform choices that depend on budget. By detailing these barriers with practical examples, this paper aims to help educators and academic leaders develop more effective strategies to adopt low-power IoT in engineering programs.
\end{abstract}


\section{Introduction 
}
\label{sec:introduction}



In today’s interconnected world, digital tools link different devices to perform a diverse range of tasks. Their use must be examined based on the suitability of the application, the context of deployment, and the task itself. In engineering education, short- and long-range IoT technologies can be adopted not only for their technical reliability, but also to build practical skills in students and future technicians \cite{iqbal2023review, Fuertes2021}.

Transmission technologies continue to evolve, driven by the need for secure connections and user requirements such as data rate, coverage, frequency, and energy consumption \cite{uzoka2024role}. These technologies appear in many settings: short-range deployments in laboratories or classrooms and long-range deployments in areas such as agriculture or environmental monitoring \cite{Negm2022, Fuertes2021}. Among these, battery life remains a major concern, influencing network design and feasibility \cite{KRISHNA2024102770, callebaut2021art}.

Low-power IoT technologies offer solutions for these challenges by balancing performance with energy efficiency \cite{Mao2021IIoT}. However, their adoption also depends on financial feasibility and the realities of academic settings, including security issues, privacy policies, reliable network connectivity, interoperability between vendors, and scalability to mention a few \cite{Madni2022-Frontiers}. These challenges can affect engineering students' exposure to industry practices, creating an imbalance between what academia can teach and what industry expects from new graduates \cite{Madni2022-Frontiers, peixoto2020analyzing}.

This paper surveys the literature on the barriers that limit the adoption of low-power IoT in engineering programs. We organize the evidence from recent studies and practical cases and focus the analysis on three barrier groupings: technical, organizational, and curricular/pedagogical. Our survey is guided by the research question (RQ): 

\textit{RQ: What barriers limit the integration of low-power IoT in engineering curricula?}




The remainder of this paper is organized as follows: Section \ref{sec:Methodology} explains the methodological considerations. Section \ref{sec:snapshot} reviews the main technical features of selected short- and long-range IoT technologies. Section \ref{sec:barriers} discusses barriers to adoption within academia. 
Finally, Section \ref{sec:conclusions} presents conclusions and future directions.

\section{Methodology
}
\label{sec:Methodology}


This study follows a thematic approach, combining a survey of the literature and selected case studies to explore barriers to IoT adoption and practical situations that strengthen collaboration between academia and industry. A survey review format was chosen because it allows a structured presentation of findings with state-of-the-art technical information that aligns with the research question \cite{mcnabb2019write}.

Although our study is guided by our RQ, it does not follow a PRISMA protocol or the exhaustive screening typical of systematic literature reviews \cite{grant2009typology}. This approach is intentional because the goal is to provide a broad, practice-oriented overview that captures both technical aspects and practical adoption experiences, rather than a purely systematic synthesis. The main methodological considerations are listed in Table \ref{tab:method-considerations}.

\section{Low-Power IoT Snapshot for Educators and Practitioners
}
\label{sec:snapshot}



Short-range and long-range technologies serve different purposes, and they can also be combined depending on the goals of each academic or practical field. For example, short-range technologies can be a suitable option for automatic control in agriculture, buildings, or small-scale industries \cite{Effah2024Agri-IoT, baidal2020design, rondon2017evaluating}. In these spaces, the short distance between transmitters and receivers makes it easier to perform automated tasks. In contrast, long-range technologies are more beneficial for monitoring distant or hard-to-reach areas \cite{GuhaRoy2025-NB-IoT, Alves2025LATAM, jiawey2024lorawan}. These situations require strong and reliable systems to track environmental conditions, habitats, or possible risks.

\begin{table}[t]
  \centering
  \caption{Main methodological considerations}
  \label{tab:method-considerations}
  \small
  \renewcommand{\arraystretch}{1.15}
  \begin{tabularx}{\columnwidth}{@{}l X@{}}
    \toprule
    \textbf{Item} & \textbf{Description} \\
    \midrule
    Approach &
    Thematic, non-PRISMA literature survey with selected case studies; flexible to capture emerging trends.
    \\
    Scope &
    Works and case reports that show barriers to adopting low-power IoT in higher education (technical, organizational, curricular/pedagogical), plus short, practice-oriented examples from academic–industry contexts.  \\
    Sources &
    Articles indexed on Scopus, Web of Science, Engineering Village. Included only peer-reviewed work aligned with the RQ and exclude sources that fall outside the scope. \\
    Keywords &
    Low-power IoT (BLE, Zigbee, Thread; LoRaWAN, NB-IoT, LTE-M), engineering curriculum, testbed, partnership, barriers, budgeting, maintenance.  \\
    Inclusion criteria &
    Publications between 2021–2025 on educational use and barriers to IoT adoption challenges; English-written papers preferred. \\
    \bottomrule
  \end{tabularx}
\end{table}

Educators and professionals need a clear understanding of how to select the right technical parameters based on what is possible with each type of technology. In academic settings, this knowledge can help connect learning outcomes to real applications, making it easier to select the right tools for both lab work and industrial applications.

\begin{figure}[!htbp]
    \includegraphics[width=0.48\textwidth]{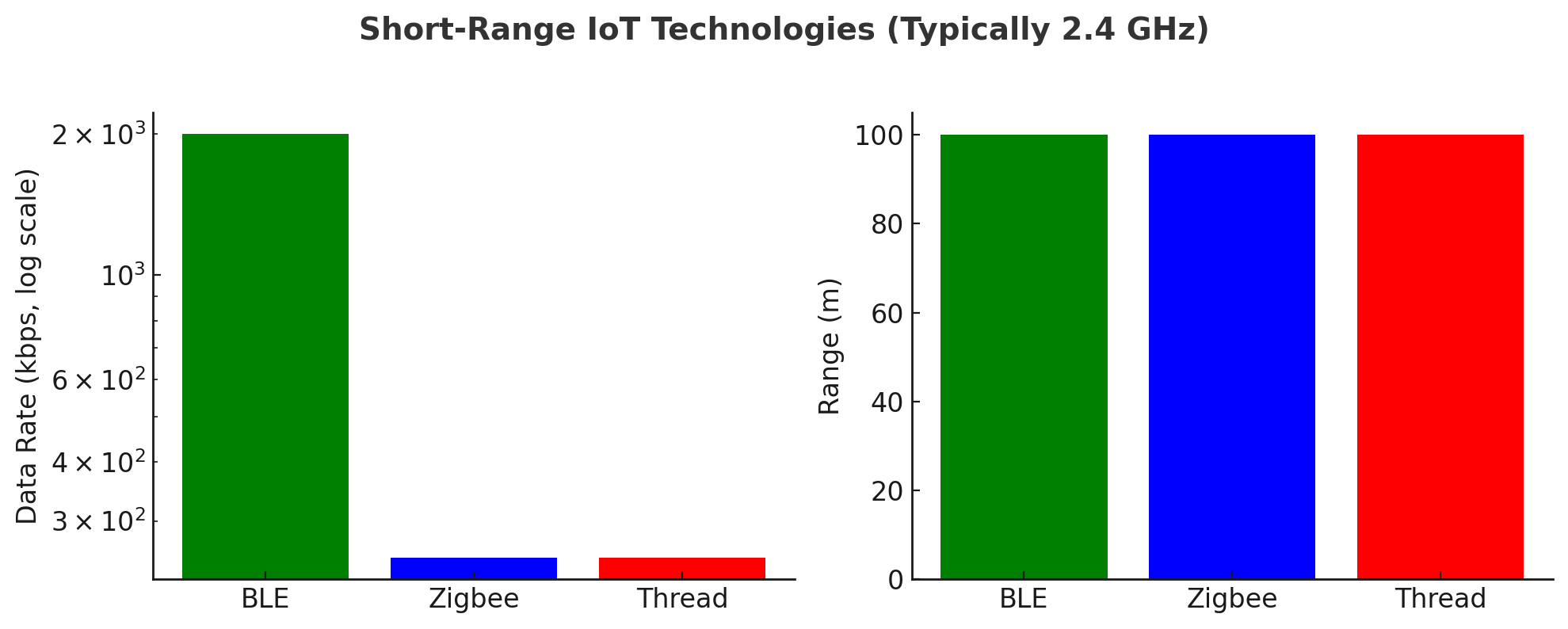}
    \caption{Short-range IoT technologies comparison}
    \label{fig:Short-Range}
\end{figure}

\begin{figure}[!htbp]
    \includegraphics[width=0.48\textwidth]{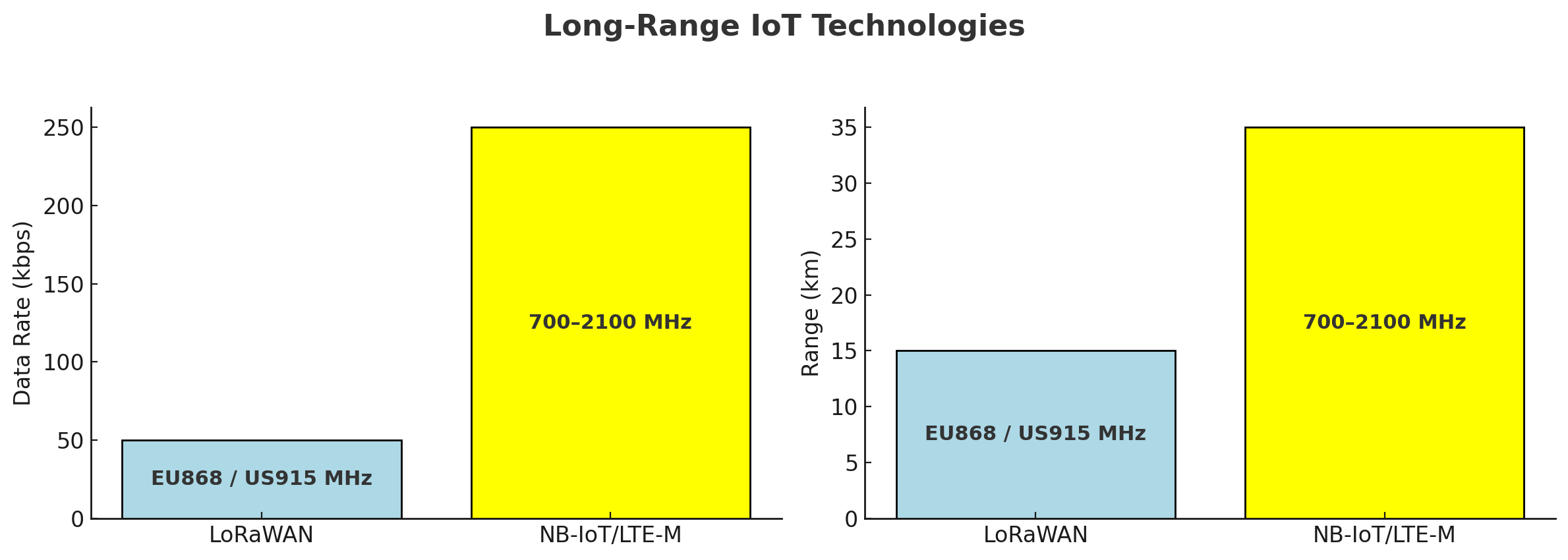}
    \caption{Long-range IoT technologies comparison}
    \label{fig:Long-Range}
\end{figure}

\begin{figure}[!htbp]
    \includegraphics[width=0.48\textwidth]{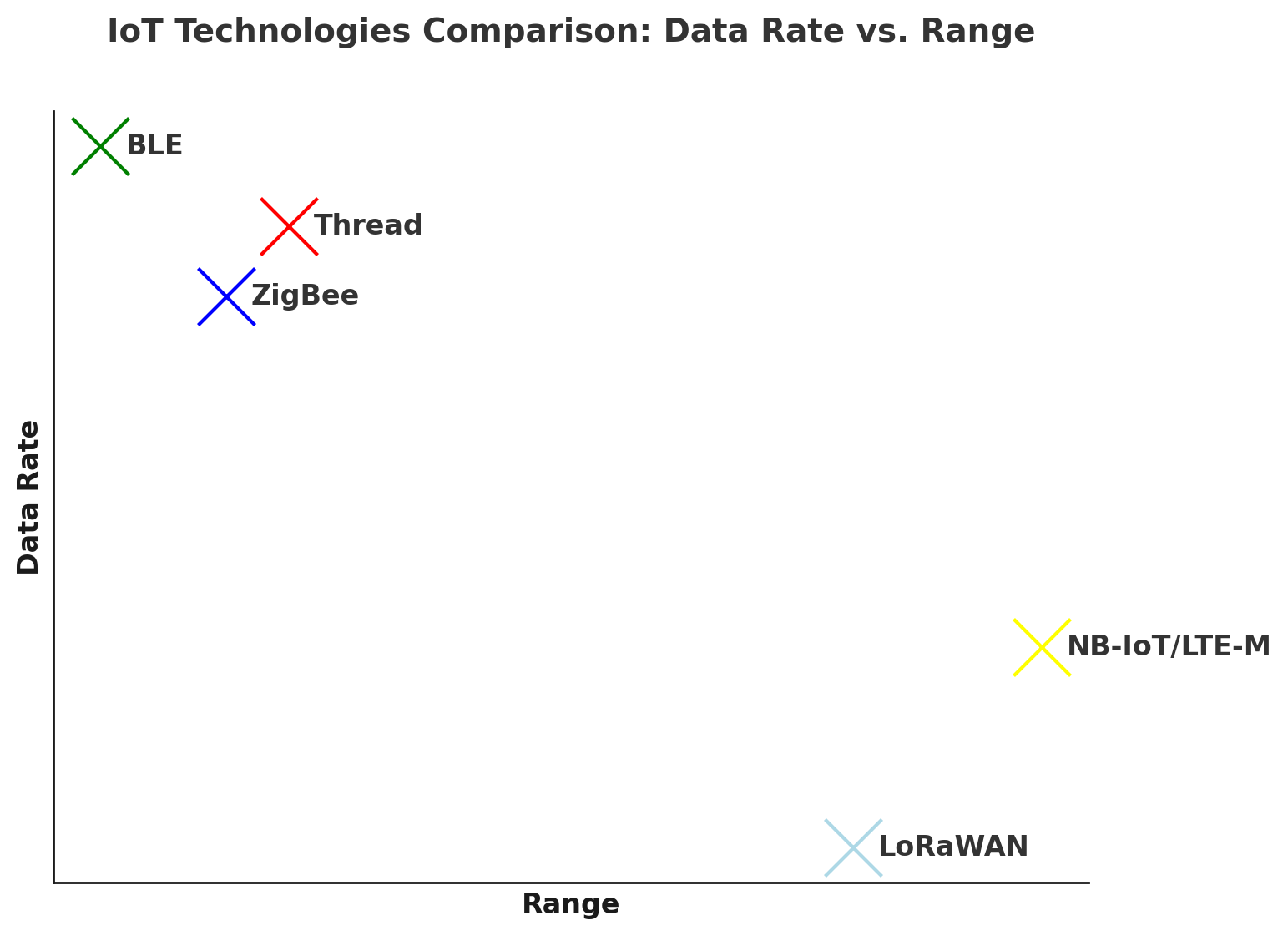}
    \caption{ Low-power IoT: Data Rate vs. Range}
    \label{fig:DataRatevsRange}
\end{figure}

The technologies shown in Figure \ref{fig:Short-Range}\cite{alobaidy2020review, KimThread2019, khattak2023performance, Nasralla2025} and Figure \ref{fig:Long-Range}\cite{Mekki2019LPWANCompare, Martinez2019NB-IoT, Adelantado2017LoRaWANLimits} are not the only examples of low-power systems. However, they were selected because they help explain how these technologies work and what kinds of needs they can meet. Table \ref{tab:low-power-edu}\cite{alobaidy2020review, KimThread2019, khattak2023performance, Nasralla2025, Ramesh2020CampusLoRaChapter, Mekki2019LPWANCompare, Martinez2019NB-IoT, Adelantado2017LoRaWANLimits, Raza2017LPWANOverview, Centenaro2016RisingStars} presents technical and cost parameters based on approximate values, summarized from several public technical documents. The goal is to offer a comparative view using specific engineering parameters and values. These are theoretical references, and real-world performance may change depending on indoor or outdoor conditions. For example, LoRaWAN and NB-IoT/LTE-M can reach up to around 15 km and 35 km in rural areas, respectively. Nevertheless, in urban environments their coverage usually drops to about 2.5 km for LoRaWAN and between 1 to 10 km for NB-IoT/LTE-M. Figure \ref{fig:DataRatevsRange} \cite{alobaidy2020review, KimThread2019, Mekki2019LPWANCompare, khattak2023performance, Martinez2019NB-IoT, Raza2017LPWANOverview} presents a relationship between Data Rate vs. Range according to some features described in Table \ref{tab:low-power-edu} \cite{alobaidy2020review, KimThread2019, khattak2023performance, Nasralla2025, Ramesh2020CampusLoRaChapter, Mekki2019LPWANCompare, Martinez2019NB-IoT, Adelantado2017LoRaWANLimits, Raza2017LPWANOverview, Centenaro2016RisingStars}.

\begin{table*}[t]
  \centering
  \caption{Representative values for comparison of Low-Power IoT technologies \cite{alobaidy2020review, KimThread2019, khattak2023performance, Nasralla2025, Ramesh2020CampusLoRaChapter, Mekki2019LPWANCompare, Martinez2019NB-IoT, Adelantado2017LoRaWANLimits, Raza2017LPWANOverview, Centenaro2016RisingStars}}
  \label{tab:low-power-edu}
  \setlength{\tabcolsep}{6pt}
  \renewcommand{\arraystretch}{1.22}
  \begin{tabularx}{\textwidth}{lYYYYY}
  \toprule
  \textbf{Technology} &
  \textbf{Data Rate} &
  \textbf{Power Consumption} &
  \textbf{Infrastructure} &
  \textbf{Investment} &
  \textbf{Campus Feasibility} \\
  \midrule

  \multirow{2}{*}{BLE / Zigbee / Thread} &
  Up to $\sim$3\,Mbps (BLE); $\sim$250\,kbps (Zigbee/Thread) &
  Very low (battery life months–years) &
  Local hub/coordinator (e.g., SBC, 802.15.4 coordinator) &
  Low 
  &
  In-lab sensor networks and short-range exercises \\
  & \multicolumn{5}{Y}{\emph{Notes:} Typically 2.4\,GHz; BLE suits mobile/app labs; Zigbee/Thread good for mesh/IoT protocol labs.}\\[2pt]

  \multirow{2}{*}{LoRaWAN} &
  $\sim$0.3 up to 50\,kbps &
  Low (multi-year; duty-cycle dependent) &
  LoRa gateways + network server/backend &
  Moderate 
  &
  Campus-wide testbeds, coverage studies, outdoor sensing \\
  & \multicolumn{5}{Y}{\emph{Notes:} Sub-GHz ISM (e.g., EU868/US915); ADR and duty-cycle constraints; useful for city-scale coursework.}\\[2pt]

  \multirow{2}{*}{NB-IoT / LTE-M} &
  $\sim$26–250\,kbps (NB-IoT); up to $\sim$1\,Mbps (LTE-M) &
  Low–moderate (multi-year typical) &
  Cellular operator network (SIM/eSIM) &
  Moderate - High 
  &
  Field deployments via industry/telecom partners; QoS/SLA potential \\
  & \multicolumn{5}{Y}{\emph{Notes:} Licensed LTE bands (700–2100\,MHz); great for industry-sponsored capstones, smart-campus pilots.}\\

  \bottomrule
  \end{tabularx}
\end{table*}


\section{Barriers to Adoption
}
\label{sec:barriers}



\begin{figure}[!htbp]
\centering
\includegraphics[width=0.45\textwidth]{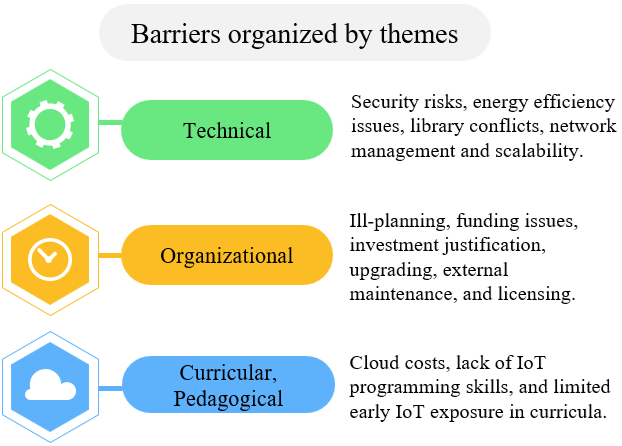}
    \caption{Barriers classified according to themes}
    \label{fig:Barriers}
\end{figure}

Transmission systems are not always ideal for adaptation or integration. Even with proper sizing, certain issues can hinder deployment or expansion, especially during startup. For low-power telecommunication systems, various barriers can emerge before full adoption. These often originate in academia, complicating industry’s ability to adopt solutions based on these technologies. The challenges identified in the literature can be categorized into technical, organizational, curricular, and pedagogical. The following subsections provide examples of how they impact the adoption of low-power technologies and limit access to scalable, reliable solutions, and other challenges in building telecommunication systems that bridge academia with industry. Figure \ref{fig:Barriers} and Table \ref{tab:barriers} summarize these barriers.

\subsection{Technical Barriers}

Technical barriers arise mainly from challenges related to infrastructure setup. They affect device management due to aspects related to energy efficiency and the ability to coexist with other wireless systems due to issues such as scalability limits, maintenance costs, and deployment challenges. These can be perceived as a challenge to overcome for educators due to investments and user-friendly platforms to tailor for educational purposes.


Infrastructure-related barriers often come from factors that increase system costs. These extra expenditures lead to more complex hardware, components, and maintenance, which can make it harder to develop educational systems due to cost overruns, especially in small lab settings. Ye \textit{et al.} \cite{Ye-AI-IoT_2021} explain that the added complexity and limitations of IoT systems, particularly when combined with AI, can raise security concerns (e.g., impractical due to the computational effort required). They also mention that adding external components, such as sensors and antennas, increases the need for more RF energy. Furthermore, using both energy harvesting and batteries makes power management more complicated, requiring efficient charging control and consideration of rechargeable batteries, which can increase system costs.

Similarly, Nurelmadina \textit{et al.} \cite{nurelmadina2021systematic} note that while industrial IoT (IIoT) networks often rely on batteries or limited power sources, which can be advantageous due to longer battery life, these systems still need high energy efficiency for operations in remote or hard-to-reach locations. They stress that choosing energy-efficient communication protocols is key to reducing power consumption and extending device lifetimes. However, they also point out that maintaining power backups, such as advanced energy harvesting or battery replacement, can be expensive and logistically difficult, especially for large-scale or remote deployments.


Researchers have illustrated the benefits of low-power IoT systems offering flexibility and scalability, mainly because of the need for less physical infrastructure and cabling, making these systems less likely to suffer cable damage caused by extreme weather or harsh environmental conditions. However, problems during sensor node deployment can still affect system performance. Jayatilake \textit{et al.} \cite{jayatilake2021design} explain that physical damage to low-power IoT components can result from power supply issues, which can cause system failures. These authors also describe how wireless sensor networks (WSNs) often use ESP microcontrollers for decentralized communication, supported by libraries that simplify system development and management. In their study \cite{jayatilake2021design}, they mention that adding extra libraries, such as ThingSpeak, can lead to conflicts at the sink nodes, making system design more difficult. These added libraries, if not working correctly, can cause irregular data transmission to the server. This issue can be reduced by using converters that support energy-efficient operation.

Low-power IoT systems also face challenges related to scalability. Aboubakar \textit{et al.} \cite{aboubakar2022review} explain that large-scale IoT deployments can make network management and scalability difficult. This becomes even more complex when the system includes different types of devices from different vendors, each with its own settings and properties. In such cases, making all devices run smoothly is not always easy. One of the main problems is that manually setting up each device is often too complicated or simply not practical. To address this, these authors suggest using self-configuring network settings that can adjust to changes such as device movement, failures, or shifts in data traffic. Even with this approach, they point out that the quality of service (QoS) can still be affected by the changing nature of the network and the limitations of low-power IoT systems.

\subsection{Organizational Barriers}

Organizational barriers can arise from issues related to procurement, readiness, and institutional culture. These barriers impact not only the costs of deploying IoT systems, but also the ongoing maintenance costs, which can be significant depending on an educational institution's budget for integrating low-power IoT into their curriculum. They can also affect faculty readiness, potentially leading to resistance if there is limited collaboration among colleagues within or across institutions.

Ensuring the long-term operation and support of IoT systems is especially difficult when funding and expertise are limited. Madni \textit{et al.} \cite{Madni2022-Frontiers} emphasize that institutional commitment, particularly financial, is essential for supporting IoT education, including e-learning. This is because the adoption of IoT requires tools, software, and specialized personnel. Effective planning and budgeting are crucial for using resources efficiently. The authors also highlight the need for institutions to understand the processes that support smooth technology adoption, and for faculty to be trained to teach and design courses around emerging technologies like IoT.

Stornelli \textit{et al.} \cite{STORNELLI2021104229} note that adopting technologies related to manufacturing and IoT involves IT infrastructure costs and expenses for adapting existing systems. Institutions often face challenges in justifying investments in infrastructure and platform integration. Furthermore, relationships with vendors can present obstacles, such as contractual issues or a lack of trust, creating uncertainty about whether technological needs will be met. In their study, they also note that the institutional culture must align with the demands of technology adoption. Without this alignment, resistance to digital change may increase, limiting the preparation for Industry 4.0 \cite{STORNELLI2021104229}.

From a broader IoT perspective, Waqar \textit{et al.} \cite{WAQAR2024e31035} highlight concerns around data security and privacy, where data breaches can become a significant issue in campus or community-wide sensor networks. They also point out that upgrading legacy infrastructure is often difficult, making it difficult to integrate new systems. Standardization adds another layer of complexity, as it can hinder interoperability between devices from different vendors. These challenges can lead to skepticism and reluctance to embrace IoT, compromising its long-term potential \cite{WAQAR2024e31035}.

Following a practical perspective, López \textit{et al.} \cite{OnelLopez2023} point out that maintenance is a key barrier to the adoption of low-power IoT devices. These devices often rely on external maintenance, such as battery replacement or firmware upgrades. Such tasks can increase costs due to the need for operational support to reduce downtime and ensure scalability, particularly in industrial or environmental monitoring networks. The authors suggest focusing on more sustainable IoT ecosystems that integrate energy harvesting. This approach can improve power management and extend the lifecycles of devices \cite{OnelLopez2023}.

Barriers can also be structural, driven by resource limits and institutional policies. Bollineni \textit{et al.} \cite{Bollineni2025} explore this issue from the perspective of healthcare systems, highlighting software licensing, data storage, and cybersecurity as cost factors for IoT networks. The authors associate these costs to proprietary datasets and platform licensing from the provider side. This view can be applied to academic settings, where similar challenges can create obstacles in training personnel. If lower-cost options (e.g.,open-source tools) are not considered, engineering graduates may face skill gaps. These barriers can also limit collaboration between universities and industry by slowing down technology transfer.

\subsection{Curricular and Pedagogical Barriers}

Curricular barriers often arise from a range of sources or a combination of them, including financial limitations. These barriers can affect the inclusion of IoT in engineering programs. Students may have limited access to cutting-edge software or hardware (open-source or proprietary) in theoretical or laboratory instructional settings which can affect their hands-on experience. This lack of practice can make it more difficult for students to advance their skills needed for industry and research positions.

Additionally, integration gaps can lead to a mismatch in learning objectives. For example, Abichandani \textit{et al.} \cite{Abichandani2022} point out that when adopting IoT systems in education, especially those involving cloud-based platforms, institutions need to consider the cost of implementation. The choice of connectivity for data collection and analysis also matters, particularly when using low-power technologies. Zigbee and Bluetooth seem to be less commonly used than Wi-Fi or Ethernet \cite{Abichandani2022}. These choices can create additional maintenance works for engineering programs and influence how testbeds or labs are set up for effective IoT training.

\begin{table*}[!t]
  \centering
  \caption{Summary of barriers to adopt low-power IoT in engineering education}
  \label{tab:barriers}
  \footnotesize                       
  \setlength{\tabcolsep}{4pt}         
  \renewcommand{\arraystretch}{1.08}  
  \begin{tabularx}{\textwidth}{@{}l l l Y@{}}
    \toprule
    \textbf{Type} & \textbf{Source} & \textbf{Technologies / Settings} & \textbf{Main issue / Brief description} \\
    \midrule
    \multirow{4}{*}{\textbf{Technical}}
      & Ye et al.\,\cite{Ye-AI-IoT_2021}          & AIoT / low-power IoT         & Power/energy management; battery vs. harvesting increases cost and design effort. \\
      & Nurelmadina et al.\,\cite{nurelmadina2021systematic} & LPWAN (cognitive-radio IIoT) & Energy efficiency and backups; remote nodes raise maintenance cost. \\
      & Jayatilake et al.\,\cite{jayatilake2021design}  & Wi-Fi WSN (ESP + ThingSpeak) & Library conflicts at sink; power faults damaging nodes; unstable data until mitigated. \\
      & Aboubakar et al.\,\cite{aboubakar2022review}   & Multi-vendor IoT management       & Scalability/QoS drops with heterogeneity and manual setup; need self-configuring management. \\
    \midrule
    \multirow{5}{*}{\textbf{Organizational}}
      & Madni et al.\,\cite{Madni2022-Frontiers}        & Education context            & Funding, planning, and staff training required for sustained deployments. \\
      & Stornelli et al.\,\cite{STORNELLI2021104229}   & Industry 4.0 / IoT           & IT/integration cost; vendor alignment and culture slow investment. \\
      & Waqar et al.\,\cite{WAQAR2024e31035}       & Smart buildings / IoT        & Security, privacy, legacy systems; interoperability/trust delay adoption. \\
      & López et al.\,\cite{OnelLopez2023}       & Energy-sustainable IoT       & Battery recharging and maintenance, firmware updates drive expenditures; energy harvesting and sustainability aspects to extend lifecycle. \\
      & Bollineni et al.\,\cite{Bollineni2025}   & Healthcare IoT               & Licensing, storage, cybersecurity; proprietary datasets limit affordable training. \\
    \midrule
    \multirow{3}{*}{\textbf{Curricular/Pedagogical}}
      & Abichandani et al.\,\cite{Abichandani2022} & BLE/Zigbee; Wi-Fi/Ethernet   & Platform/cloud cost shapes lab design; connectivity choices affect maintenance. \\
      & Lagogiannis et al.\,\cite{lagogiannis2025no} & Bluetooth (chosen); LoRa/Wi-Fi & No-code for non-programmers; Bluetooth fits budget but limits range/scalability. \\
      & Hachani \& Ajailia\,\cite{Hachani2023-Civil} & ESP32/Arduino + ThingSpeak   & Non-ECE students short on ICT; PBL with low-code improves readiness and practice. \\
    \bottomrule
  \end{tabularx}
\end{table*}

Some researchers can perceive a typical IoT-focused course as a challenge when assuming that students already have programming experience. According to this scenario, Lagogiannis \& Chatzopoulos \cite{lagogiannis2025no} designed a course for students with no background in programming. They also mention that some engineering programs do not include any programming-focused courses, limiting students' exposure. Based on their experience, many no-code IoT platforms rely on connectivity options that are not entirely free, such as WiFi or LoRa. LoRa, being long-range and low-power, was not needed for their educational goals. These situations can show a lack of flexible integration options, which may make certain technologies too costly to adopt in a curriculum. Yet, for their course design, the authors chose Bluetooth as an alternative for their tests due to its short-range scope, making it suitable for laboratory activities and keeping costs manageable, although they note its lack of scalability as a limitation.

Pedagogical practices also shape IoT curriculum development, even in programs outside the field of electrical and computer engineering. To address poor teaching approaches, researchers explore options like Project-Based Learning (PBL) and low-to-no-code tools. Hachani \& Ajailia \cite{Hachani2023-Civil} argued the challenge of teaching IoT to civil engineering students with limited ICT experience, especially given tight teaching schedules and little background in embedded or mobile systems. Their study shows that students value early exposure to ICT and IoT concepts in the curriculum. They also report that using PBL, along with the ESP32 and Arduino IDE (a low-code tool), and ThingSpeak for cloud integration, helped students become better prepared to analyze requirements in civil engineering projects.







\section{Conclusion}
\label{sec:conclusions}


This paper examines barriers to the adoption of IoT in academia and highlights how this adoption can shape interaction with environments that rely on practice-oriented learning, which is an essential factor for societies aiming to advance technologically in the era of digital and smart solutions. These barriers fall into three broad categories: technical, organizational, and curricular/pedagogical. Although there are alternative ways these barriers can be grouped, we present these categories as a generic classification according to applicable literature. The challenges in adoption of IoT include power constraints, licensing and dataset costs, and the need to introduce IoT concepts early in engineering programs.

We highlight the ways in which curricular gaps, limited programming readiness, and high implementation costs can create significant obstacles to integrating IoT in engineering education, especially when aiming to align with STEM standards. Our findings also point to the need for short- and long-term strategies to incorporate smart solutions into the classroom. Scaling these solutions, however, remains a key challenge, often affecting how projects are developed and applied in teaching environments. In addition, supporting early student exposure to IoT and applied technologies is essential to improving engineering education and research. Future work can explore ways to overcome these barriers by focusing on partnerships, funding, and access to tools and data for students and institutions.



\end{document}